\documentclass[floats,floatfix,showpacs,amssymb,prl,twocolumn,superscriptaddress,nofootinbib,longbibliography]{revtex4-1}

\usepackage[normalem]{ulem}
\usepackage{booktabs}
\usepackage{graphicx}%
\usepackage{dcolumn}%
\usepackage{bm}%
\usepackage{amsmath}
\usepackage{xcolor}
\usepackage{xspace}
\usepackage[utf8]{inputenc}
\usepackage[T1]{fontenc}

\newcommand{\dl}{luminosity distance\xspace}

\newcommand{\msun}{M$_\odot$\xspace}
\newcommand{\hnot}{\ensuremath{\mathrm{H}_0}\xspace}
\newcommand{\tjn}{\ensuremath{\theta_\mathrm{JN}}\xspace}
\newcommand{\degc}{\ensuremath{\circ}\xspace}

\newcommand{\va}{viewing angle\xspace}
\newcommand{\sd}{Schutz distribution\xspace}
\newcommand{\tjet}{\ensuremath{\theta_{\rm up}}\xspace}

\newcommand{\linf}{\textsc{LALInference}\xspace}
\newcommand{\nevents}{\ensuremath{4}\xspace}

\newcommand{\firstevent}{B\xspace}
\newcommand{\polarevent}{C\xspace}
\newcommand{\loudevent}{A\xspace}
\newcommand{\weakevent}{D\xspace}

\newif\ifthetajn \thetajnfalse

\def\mnras{{Mon. Not. R. Astron. Soc.}}             
\def\araa{{Annu. Rev. Aston. Astrophys.}}             
\def\apjl{{Astrophys. J. Lett.}}                
\def\aap{{Astron. Astrophys.}}                
\def\nar{\ref@jnl{New A Rev.}}          
\def\nat{Nature}               

\begin{document}

\title{Viewing angle of binary neutron star mergers}
\author{Hsin-Yu Chen}
\email{hsinyuchen@fas.harvard.edu}
\affiliation{Black Hole Initiative, Harvard University, Cambridge, Massachusetts 02138, USA}
\author{Salvatore Vitale}
\email{salvatore.vitale@ligo.mit.edu}
\affiliation{LIGO Laboratory and Kavli Institute for Astrophysics and Space Research, Massachusetts Institute of Technology, Cambridge, Massachusetts 02139, USA}
\author{Ramesh Narayan}
\email{rnarayan@cfa.harvard.edu}
\affiliation{Harvard-Smithsonian Center for Astrophysics, Harvard University, Cambridge, Massachusetts 02138, USA}

\date{\today}

\begin{abstract}
The joint detection of the gravitational wave (GW) GW170817 and its electromagnetic (EM) counterparts 
GRB170817A and kilonova AT 2017gfo has triggered extensive study of the EM emission of binary neutron star mergers. 
A parameter which is common to and plays a key role in both the GW  and the EM analyses is the viewing angle of the binary's orbit.
If a binary is viewed from different angles, the amount of GW energy changes (implying that orientation and distance are correlated) and 
the EM signatures can vary, depending on the structure of the emission. Information about the viewing angle of the binary orbital plane is therefore crucial to the interpretation of both the GW and the EM data, and can potentially be extracted from either side.

In the first part of this study, we present a systematic analysis of how well the viewing angle of binary neutron stars can be measured from the GW data.
We show that if the sky position and the redshift of the binary can be identified via the EM counterpart and an associated host galaxy, then for 50$\%$ of the systems the viewing angle can be constrained to $\leq 7^{\circ}$ uncertainty from the GW data, independent of electromagnetic emission models. 
On the other hand, if no redshift measurement is available, the measurement of the viewing angle with GW alone is 
not informative, unless the true viewing angle is close to $90^\degc$. This holds true even if the sky position is measured independently. 

Then, we consider the case where some constraints on the viewing angle can be placed from the EM data itself. We show that the EM measurements 
can then be used in the analysis of GW data to improve the precision of the luminosity distance, and hence of the Hubble constant, by a factor of 2 to 3. 
 
\end{abstract}

\maketitle

\section{Introduction}

The discovery of binary neutron star (BNS) merger GW170817~\citep{2017PhRvL.119p1101A} 
and its electromagnetic (EM) counterparts, GRB170817A and kilonova AT 2017gfo, opened the era of multi-messenger astronomy~\citep{2017ApJ...848L..12A,2017ApJ...848L..13A,2017Sci...358.1556C,2017ApJ...848L..16S}. 

While much was learned from this first joint detection, the precise EM emission model is still unknown. 
A key parameter to understand the EM emission is the orbital inclination 
angle~\footnote{In the gravitational-wave literature, the inclination angle is usually the angle between the line of sight 
and the \emph{orbital} angular momentum. However, following LIGO and Virgo, we report the angle between the line of sight 
and the \emph{total} angular momentum. The difference is only important for large and misaligned spins, which are 
not expected for binary neutron star mergers.}, which strongly impacts the details of the EM signals received at Earth.
{For example, t}he gamma-ray burst (GRB) GRB170817A was under-luminous~\citep{2017ApJ...848L..13A,2017ApJ...848L..14G}, 
and the X-ray and radio emission that followed were significantly different from those of other GRBs~\citep{2014PhRvD..90b4060A,2017Natur.551...71T,2017ApJ...848L..20M,2017Sci...358.1579H,2017ApJ...848L..21A}.
A possible explanation could be that a GRB jet was formed, which we observed from a large inclination angle. 
However, alternative explanations exist. The observations could also be described by a choked jet, whose cocoon expands leading to a wide-angle, mildly relativistic outflow~\citep{2017Sci...358.1559K,2017arXiv171005896G}.
If the choked jet is the correct scenario, the inclination angle affects less the light curve and the spectra, and thus cannot be constrained by the available EM data.
{Although recent follow-up observations seem to favor  the off-axis emission explanation 
for GW170817/GRB170817A~\citep{2018arXiv180609693M,2018arXiv180806617V,2018arXiv180800469G}, 
 other models have not definitively been ruled out, and will be tested with future joint detections.} The inclination angle is also a key ingredient for the interpretation of 
the kilonova emission and its spectral evolution~\citep{2015MNRAS.450.1777K,2017ApJ...848L..17C,2017ApJ...851L..21V,2017Natur.551...71T,2017ApJ...848L..27T}. 

It would thus be important if the inclination angle of the binary, \tjn, could be measured or at least constrained from the GW data, 
and then used to study or exclude EM emission mechanisms.

Unfortunately, measurement of the inclination angle with GW data is usually quite poor, due to the 
well-known degeneracy between the inclination and 
the \dl ~\citep{2016PhRvL.116x1102A}.
This degeneracy can be resolved if the system has precessing spins, if higher-order harmonics are detectable, or if the merger and ringdown are in band. 
None of these conditions are met for BNS, since neutron stars have small spins, and mass ratios close to one, which suppresses the 
amplitude of higher-order harmonics~\citep{PhysRevD.49.6274,2004PhRvD..70d2001V,2014PhRvL.112y1101V,2017PhRvD..95f4053V,2015PhRvD..92b2002G,2018PhRvL.120p1102L,2018arXiv180611160C}.
However, in the case of a joint GW-EM detection more information usually exists. 
{If an independent measurement of the luminosity distance can be made from the redshift of the EM counterpart, or using other properties of the host~\citep{2018ApJ...854L..31C}},
the degeneracy is broken and one can expect the uncertainty on the inclination angle to improve.
To a smaller extent, knowledge of sky position can also help to improve the measurement of the inclination angle, as we discuss below.
One could thus envisage a strategy where some information obtained from the EM {analysis} (redshift, sky position) can be folded into the GW analysis to get an improved measurement of the inclination angle.
{ We note that although GW detectors are able to measure the binary \emph{inclination angle} $\tjn$, which also carries directional 
information on the binary rotation (clockwise or counterclockwise) and has a range between $0^{\degc}$ and $180^{\degc}$, most EM observations 
only depend on the binary \emph{\va} $\zeta$, which is defined as $\zeta\equiv \mathrm{min}(\tjn, 180^\circ - \tjn)$~\cite{2017ApJ...848L..13A}. 
We hereby focus on the \va in this paper. Using} the sky position of the kilonova AT 2017gfo and the redshift of the host galaxy NGC4993, 
the bound on the \va of GW170817 was improved from $\zeta \lesssim 56^{\circ}$~\citep{2017PhRvL.119p1101A} to $\zeta \lesssim 28^{\circ}$~\citep{2017PhRvL.119p1101A,2018arXiv180511579T,2018ApJ...860L...2F}.

In the future one might also consider the opposite approach: if the details of the EM emission are well understood, the detection 
of photons could provide a bound on the \va.
For example, detection of a short GRB with a jet break in the afterglow can yield an upper-bound on the binary \va. 
{The jet break indicates that we observed the short GRB within 
the core of its relativistic jet~\citep{2006ApJ...653..468B,2013PhRvL.111r1101C,2014ARA&A..52...43B,2015ApJ...815..102F}. 
If the jet is aligned with the binary rotation axis, the jet break constrains the maximum viewing angle of the binary. 
In addition, the observations of GRB170817A and the kilonova AT 2017gfo have led to more thorough studies on the EM emission models, 
providing measurements of GW170817's viewing angle even without the detection of a jet break (e.g.~\citep{2017ApJ...848L..20M,2018arXiv180609693M,2018arXiv180906843W}).
This information} can be used in the GW analysis to get a better measurement of the \dl~\citep{2017PhRvL.119r1102F}, 
which in turn can improve the Hubble constant measurement with GWs~\citep{2017ApJ...851L..36G,2018arXiv180610596H}.  

In this paper we present a systematic study of the measurability of the \va of BNS systems. 
We first show that, in the absence of a positive detection of an EM counterpart, GWs alone only rarely provide a meaningful constraint on the \va.
We then consider the case when the \dl and/or the sky position are independently measured. While little is changed by knowing the sky position, 
knowledge of the \dl dramatically reduces the $1\sigma$ uncertainty for the viewing angle. For $50\%$ of the BNS detections for which sky position and \dl are 
measured from  EM {data}, the viewing angle uncertainty is below  $7^{\circ}$ (using the projected sensitivities for LIGO and Virgo in the third science run). 
 {This uncertainty is small enough to allow for interesting comparison to EM emission models.}

{In this paper we present a systematic study of the measurement of GW binary \va using auxiliary EM information. 
The end-to-end set of simulations we perform was not previously possible due to the 
expensive computational costs required to carry fully numerical MCMC parameter estimation runs for a large number of sources~\cite{2015PhRvD..91d2003V} . 
We have circumvented this issue by running the full MCMC algorithm~\cite{2015PhRvD..91d2003V} for a subset of sources and shown how a different and faster 
dedicated code we have written yields the same results. This faster code is what allows us to carry out the study we describe in this paper.}

We also consider the opposite scenario, and show that if the binary viewing angle is constrained by the EM data, the binary \dl uncertainty can potentially be reduced by a factor of 2 to 3. In turn, 
that improves the measurement of the Hubble constant with GWs.
We show that a $1\%$ uncertainty on the Hubble constant can then be reached with only $O(10)$ GW-EM joint detections of BNS, that is, in less than 5 years.

{We note that Refs.~\citep{2006PhRvD..74f3006D,2010ApJ...725..496N,2017PhRvL.119r1102F} have discussed 
the improvement on the Hubble constant measurement with GWs when a short GRB counterpart is found. 
In these studies, the jet is assumed to be aligned with the binary rotation axis, and 
the short GRB observations provide upper-bounds for the binary viewing angle
(in the form of a top-hat prior or some other  probability distribution, usually with a maximum at $\zeta=0^{\degc}$, and decreasing at higher angles).
In one of the analyses we carry out, we consider different values for the upper-bound on the \va from EM observations. 
However, we find that the resulting improvement in the distance measurement does not strongly 
depend on how tight the upper-bound is, as long as the bound is below $\sim 30^{\degc}$. 
In addition to this upper-bound approach, we also consider a new and more realistic scenario where 
the binary \va is not constrained from $0^{\degc}$ to some upper-bound, but rather is given as a normal distribution 
centered at different values (from $0^{\degc}$ to $90^{\degc}$). 
This extends what was done in previous literature, which exclusively focused on small viewing angles.
We argue that it is important to not only focus on small \va, since future events will usually have moderate viewing angles ($\sim 30^\degc$)~\citep{2011CQGra..28l5023S}.} 
{Even with a \va as large as $30^\degc$, EM observations can still yield meaningful constraints, as shown by the 
follow-up of GW170817.

\section{Method and Results}

We consider two different methods to estimate the distance and inclination of BNS systems. 
First, we rely on the computationally-expensive stochastic sampler \linf~\cite{2015PhRvD..91d2003V} (specifically on its nested sampling flavor~\cite{2010PhRvD..81f2003V}).
This is the same algorithm used by the LIGO and Virgo collaborations and delivers posterior distributions for all the 
unknown parameters on which compact binaries depend.
Given the cost of each simulation~\cite{2015PhRvD..91d2003V}, \linf cannot be run on an arbitrarily large number of simulations. We thus use  it only on a few specific sources, to show which parameters affect the measurability of \dl and inclination, and how.
We then introduce a semi-analytical, faster, approach. After showing that the two give consistent results we use 
the latter to characterize the population of detectable BNSs.
 
\subsection{Single-event analysis}

\begin{table}
\begin{tabular}{lllllll}
\hline\hline
  & SNR  & polarization  & phase & RA & DEC   \\
  \hline\hline
Source \loudevent & 35 &0.005&0&-1.08 & 0.66 \\ 
Source \firstevent & 20 &0.005&0&-1.08 & 0.66\\ 
Source \polarevent & 20 &0.005&0& 0 & $\pi$/2 \\ 
Source \weakevent & 12 &0.017 &0.017&-1.08 & 0.66 \\ 
\hline \hline
\end{tabular}
\addtolength{\tabcolsep}{-10pt}
\caption{For all events the GPS time is 1068936994.0 and the two component masses in the source frame are $1.4$\msun.}\label{Tab.injs}
\end{table}

All the BNS systems we simulate have component masses $1.4-1.4$\msun in the \emph{source frame}. As mass measured in the detector 
frame is redshifted by a factor $(1+z)$, where $z$ is the redshift, the sources will appear slightly heavier in the \emph{detector frame}.

To keep the computational cost reasonable, we make two main simplifying assumptions: we neglect tidal effects and neutron star spins.
The former is a very reasonable choice, since tidal effects do not enter the waveform amplitude, and hence are not correlated with the inclination angle. 
The latter is justified since the spins of known neutron stars in binaries that will merge within a Hubble time is small 
(the fastest-spinning systems are PSR J0737-3039A~\citep{2003Natur.426..531B} and PSR J1946+2052~\citep{2018ApJ...854L..22S}, 
which will at most have dimensionless spins of $\chi \sim 0.04$ or $\chi \sim 0.05$ when they merge). 
Even GW170817 is consistent with having small spins~\citep{2018arXiv180511579T}. 

In the work reported here, all the synthetic BNS signals are generated using the IMRPhenomPv2 waveform 
family~\cite{2015PhRvD..91b4043S,Hannam:2013oca}, with the reduced order quadrature likelihood approximation~\cite{Smith:2016qas}. 
We consider a network consisting of the two LIGOs~\citep{2015CQGra..32g4001L} and Virgo~\citep{2015CQGra..32b4001A}, all at design sensitivity. 
The signals are added into `zero-noise' (which yields the same results that would be obtained averaging over many noise realizations). 
We start all analyses at $20$~Hz, and use a sampling rate of $8$~KHz. We marginalize over calibration errors using the same method described in Ref~\cite{GW150914-PARAMESTIM}, using gaussian priors with widths of 3\% for the amplitude and 1.5 degrees phase for all instruments.

We consider two different sky positions, to verify if and to what extent our conclusions depend on the 
detector antenna patterns. In total we create \nevents such systems. Their parameters are summarized in Table~\ref{Tab.injs}. 
For sources \loudevent, \firstevent and \weakevent, the sky position is near the maximum of LIGO's antenna pattern, where one would expect most detections to be made~\cite{2017ApJ...835...31C}. 

The main goal of the present study is to verify how well the inclination angle can be measured. Obviously, this would a priori depend on 
three main factors: the SNR of the event, the true value of the inclination angle, and the sky location of the event.
Each of the sources listed in Table~\ref{Tab.injs} is re-analyzed for different values of the inclination angle, from nearly 
face-on ($\tjn=0^{\circ}$) to edge-on ($\tjn=90^{\circ}$). Every time the orientation angle is changed, the distance is also 
varied to keep the SNR fixed at the value given in Table~\ref{Tab.injs}.

We note that one expects most detections made by advanced detectors to have inclinations 
close to {$30^{\degc}$ or $150^{\degc}$}
, whereas events with inclination close to edge-on ($\tjn=90^{\circ}$) would be rarer~\cite{2011CQGra..28l5023S}.
Why this happens is related to the degeneracy between \dl and inclination. Since that will play a role in the interpretation of the results we present, it's worth expanding on the subject.
Geometrical arguments would suggest that the inclination angles of the population of binaries should be uniform on the sphere, 
i.e $p(\tjn) \sim \sin(\tjn)$, while their luminosity distances should be uniform in volume, $p(D_L) \sim D_L^2$ 
(as long as cosmological effects can be neglected).
However, the \emph{detected} binaries are a subset of the entire population: they are the fraction that produce a GW signal loud enough to be detected.
GW emission from a compact binary is not isotropic, instead more energy is emitted along the direction of the orbital angular momentum, while the least 
amount is emitted parallel to the orbital plane~\cite{Maggiore}.
This implies that edge-on systems, which are the most numerous in the underlying population, will need to be extremely close to be detectable. 
Conversely, face-on/off systems can be farther away, and still produce a detectable signal. But since far away there is more volume, 
the population of \emph{detectable} signals will be dominated by sources with inclination angles close to {$30^\degc$ and $150^\degc$}.
Note that all events LIGO and Virgo have detected thus far (including the binary black holes) are consistent with 
{$30^{\degc}$ or $150^{\degc}$ inclination angles}
~\cite{2016PhRvL.116x1102A,2016PhRvX...6d1015A,2017PhRvL.118v1101A,2017ApJ...851L..35A,2017PhRvL.119n1101A,2018arXiv180511579T,2018arXiv181112907T}. 
The analytical form of the inclination angle distribution for sources detectable 
by advanced detectors~\footnote{It is worth mentioning that this selection effect may be resolved as more sensitive detectors come online~\cite{2016PhRvD..94l1501V}} 
was first obtained by Ref.~\cite{2011CQGra..28l5023S}, and we will refer to it as the \sd in this work. 
Using the \sd one can calculate the fraction of detectable events that will have \va within a given range. 
In particular, less than $\sim7\%$ ($\sim3\%$) of the detectable events will have \va$>70^\circ$ ($>80^\circ$).

All analyses are performed three times: a first time assuming that {\emph{all parameters are unknown}} and measured from GW data alone; 
a second time assuming a counterpart has been found, which provides the {\emph{sky position (right ascension and declination) of the source}}; 
a third time, assuming {\emph{both sky position and distance are known}} (the latter by measuring the redshift and using a cosmology to convert redshift to luminosity distance). 

For all parameters, we use the same priors used by the LIGO and Virgo collaborations. In particular, the prior on the \dl is uniform in volume while the prior on the inclination angle is isotropic.

Since the EM emission only depends on {the \va, }
we quote results for the viewing angle $\zeta$, 
rather than the inclination angle \tjn itself.
In Figure~\ref{fig:incinc} we report the 1-$\sigma$ uncertainty (in degrees) for the measurement of the viewing angle $\zeta$
 for all sources, as a function of the true inclination angle.

We first discuss the case where no information is available from the EM side (dot-dashed lines in Figure~\ref{fig:incinc}). 
We find that the uncertainties are roughly constant until the true inclination angle gets above $\sim 80^\circ$. 
For the sources \loudevent, \firstevent and \polarevent, the uncertainty happens to be nearly the same: $\sim 15^\degc$, while it is 
a little wider for source \weakevent $\sim 19.5^\degc$. 
We have verified that for all these configurations the viewing angle posterior is not very informative. 
It is similar to the \sd, with some extra support at $30^\degc$ and a depletion of support close to-edge on.
This is shown in Figure~\ref{fig:priorCompSourceA} for Source \firstevent, and in Figures.~\ref{fig:priorCompSourceC}, ~\ref{fig:priorCompSourceD}, ~\ref{fig:priorCompSourceB} in the Appendix for the other sources.

\begin{figure}
\centering
\includegraphics[width=1\columnwidth]{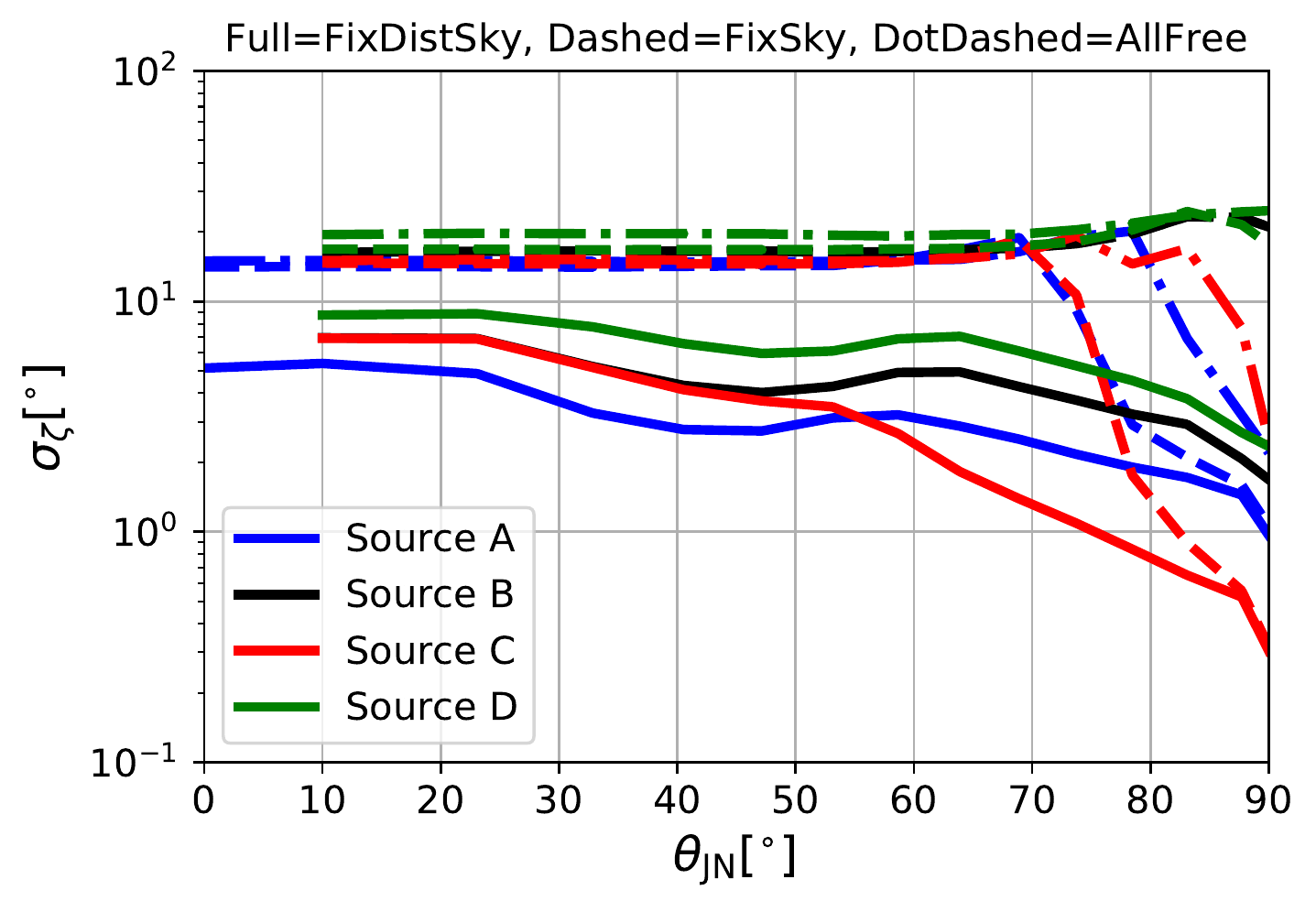}
\caption{\label{fig:incinc}
Viewing angle uncertainty as a function of binary inclination angle.}
\end{figure}

The fact that the \va uncertainty is the same for small to moderate inclinations can be explained by a combination of priors and the well-known degeneracy between \dl and inclination. 
Since the emission of GW is larger toward the direction of the system angular momentum, one can obtain a similar signal at the detector by increasing the viewing angle while moving the source closer.
As the \emph{true} inclination increases from zero, the \emph{true} distance has to decrease to maintain the same SNR. However, smaller distances are disfavored by the prior.
Thus, the Bayesian code will prefer to keep the \va posterior close to face-on and overestimate the distance to compensate. 
This can be done because of the correlation between the two parameters: the eventual small decrease in the likelihood introduced by biasing both \dl and inclination is more than compensated for by the better prior value at larger distances.
This behavior can be sustained till the true inclination angle is close to edge-on. At that point, the degeneracy is reduced, and the likelihood penalty for keeping the posterior at face-on cannot be compensated for by the prior: both \dl and \va posteriors are centered at the true value, and typically better measured~\cite{2018arXiv180407337V}. 

For the low SNR event, source \weakevent, one gets exactly the \sd for small to moderate inclinations. 
Unlike for the other sources, in this case the uncertainty \emph{increases} as the inclination angle gets close 
to $90^\circ$. This happens because when the true inclination is close to edge-on a significant posterior peak 
still survives at $30^\degc$: since the SNR is low, the extra likelihood to be gained with more support at 
edge-on is comparable with the prior penalty, and a bimodal distribution arises (see Figure~\ref{fig:priorCompSourceB} in the Appendix).

\begin{figure}
\centering
\includegraphics[width=0.95\columnwidth]{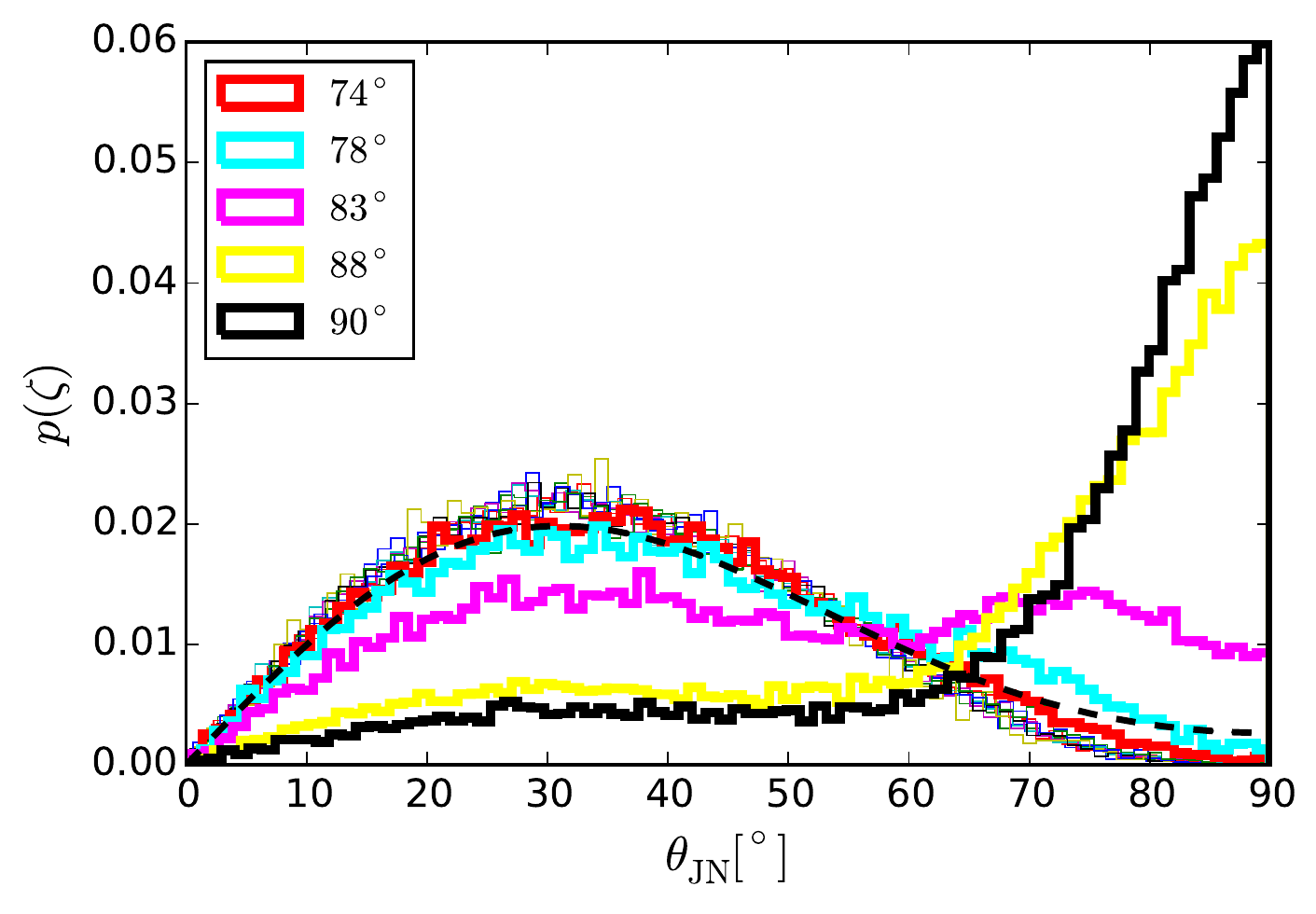}
\caption{Viewing angle posteriors for the source \firstevent, assuming no information is provided from the EM sector. The dashed line is obtained from the \sd. 
For all values of the inclination angle below $\sim 75^\circ$ the posteriors are the same (thin lines), and similar to the \sd.
We highlight posteriors for sources with large \va with thicker lines, where the value of the \va is given in the legend.}\label{fig:priorCompSourceA}
\end{figure}

We find that similar conclusions can be drawn if the sky position of the source can be considered as known (dashed lines in Figure~\ref{fig:incinc}). 
It is still the case that the posteriors for small to moderate inclinations are similar to the \sd, and 
the main difference is that the distance-inclination degeneracy is resolved at 
smaller inclinations, $\tjn \lesssim 70^\degc$ (instead of $\tjn \lesssim 80^\degc$ for the previous case). 

It is only when both sky position and distance are known (solid lines in Figure~\ref{fig:incinc}) that the uncertainties are much smaller for all sources and all inclinations.
In this case the posteriors for the \va are centered around the true value for all systems. 
The uncertainties reach a minimum when the true source is edge-on, since in that case the cross polarization of the GW signals is zero, 
which reduces the residual degeneracy between inclination and polarization angle $\psi$ (the only two unknown angles left in the 
amplitude of the GW signal).

It is interesting to verify how precisely the BNS sources can be localized as a function of their orientation. 
This is shown in Figure~\ref{fig:incarea} for the runs in which all parameters are considered unknown.
We find that the 90\% credible interval (in deg$^2$) is roughly constant for small to moderate inclination angles and for all sources.
Source \loudevent is localized better than source \weakevent because of its higher SNR. The difference between sources \firstevent and \polarevent (which have the same network SNR) is that source \polarevent has an SNR roughly split in equal amounts in the three interferometers, while source \firstevent has most of the SNR in the two LIGOs, while being sub-threshold SNR in Virgo ($4.2$ for the face-on orientation). 
Since most of the sky resolution for GW sources comes from triangulation and by requiring phase and amplitude consistency across the 
network~\citep{2011CQGra..28j5021F,2014ApJ...795..105S,2016PhRvD..93b4013S}, 
it helps if the source is above threshold in more detectors 
(Incidentally, we stress that the same is \emph{not} true for the \va uncertainty: as Figure~\ref{fig:incinc} shows one gets the same uncertainties for 
sources \firstevent and \polarevent. Likewise, the uncertainty for the \dl is the same for the two sources up to 
viewing angles of $\sim70^\circ$).
We note that the edge-on sources are relatively poorly-localized. 
This can be explained as follow: the sky position (right ascension and declination) and orientation 
(inclination, polarization) angles all enter the frequency-domain GW amplitude and phase (e.g. Eqs. 4.1-4.5 in Ref.~\cite{2011PhRvD..84j4020V}).
When the inclination angle is close to $90^\degc$, some of these terms are suppressed, reducing the number of constraints that can be used to enforce phase and amplitude consistency, leading to larger uncertainties. 

\begin{figure}
\centering
\includegraphics[width=1\columnwidth]{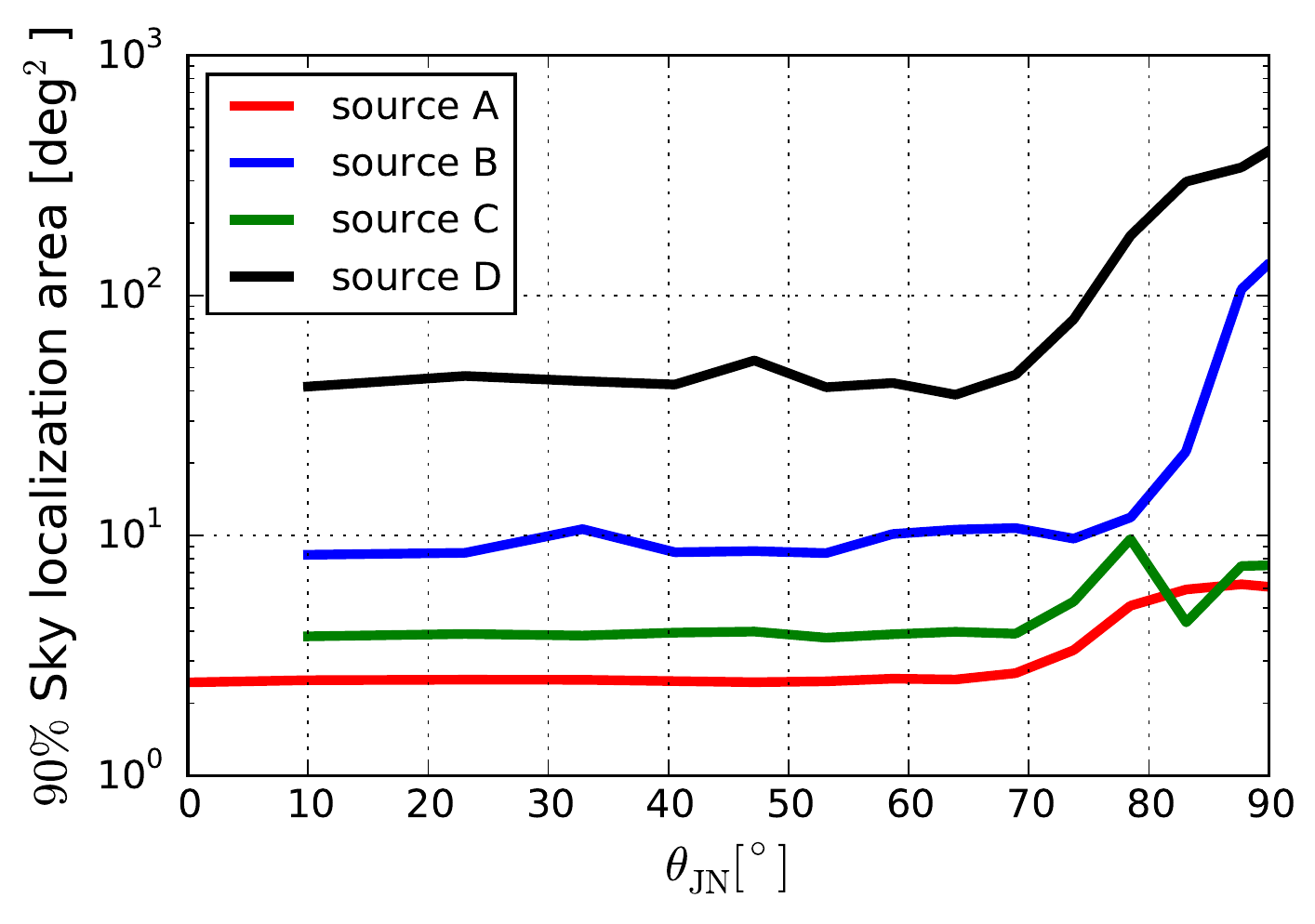}
\caption{\label{fig:incarea}
Sky localization area (in deg$^2$) as measured by \linf, as a function of binary inclination angle.}
\end{figure}

\subsection{{Computation considerations for a population analysis}}

Having gained an understanding of which parameters can impact the measurability of the \va, 
we would like to determine what fraction of the BNSs for which an EM counterpart is found will yield significant \va constraints.
Unfortunately, it is computationally prohibitive to run \linf on large sets of hundreds or thousands of events. 

We thus build an approximate Bayesian estimator for the two binary parameters of interest: the inclination angle and the \dl. 

The algorithm assumes that sky position, chirp mass, and mass ratio of the binaries are known. 
These assumptions can be justified as follows.
First, if an EM counterpart is found, it will typically provide a precise sky position. 
Second, the  mass parameters, which are inferred from the phasing of the GW waveform do not significantly couple to distance and inclination, 
which are primarily measured from the amplitude of the signal.
Since the arrival time, arrival phase (or rather, the arrival time and phase difference between detector 
pairs) and the signal-to-noise ratio are measured, 
the only unknown parameter left is the orientation of the binary in the plane  of sky, $\psi$ (``polarisation'')~\citep{2009LRR....12....2S}.
We modify the Bayesian estimator of Ref.~\cite{2015arXiv150900055C} to use the events' signal-to-noise ratios{, the relative arrival time differences }and the relative phase 
differences to reconstruct posteriors for \dl and \va, while numerically marginalizing over the polarization (see the Appendix for more details). 

We have verified that the standard deviations we obtain for the viewing angle using the approximate code are very similar to the estimates obtained with \linf for the sources described in the previous section. 
For example, in Figure~\ref{Fig.Comp} we show the uncertainties for source \weakevent when the sky position is known, 
or when both \dl and sky position are known.

\begin{figure}
\centering
\includegraphics[width=1\columnwidth]{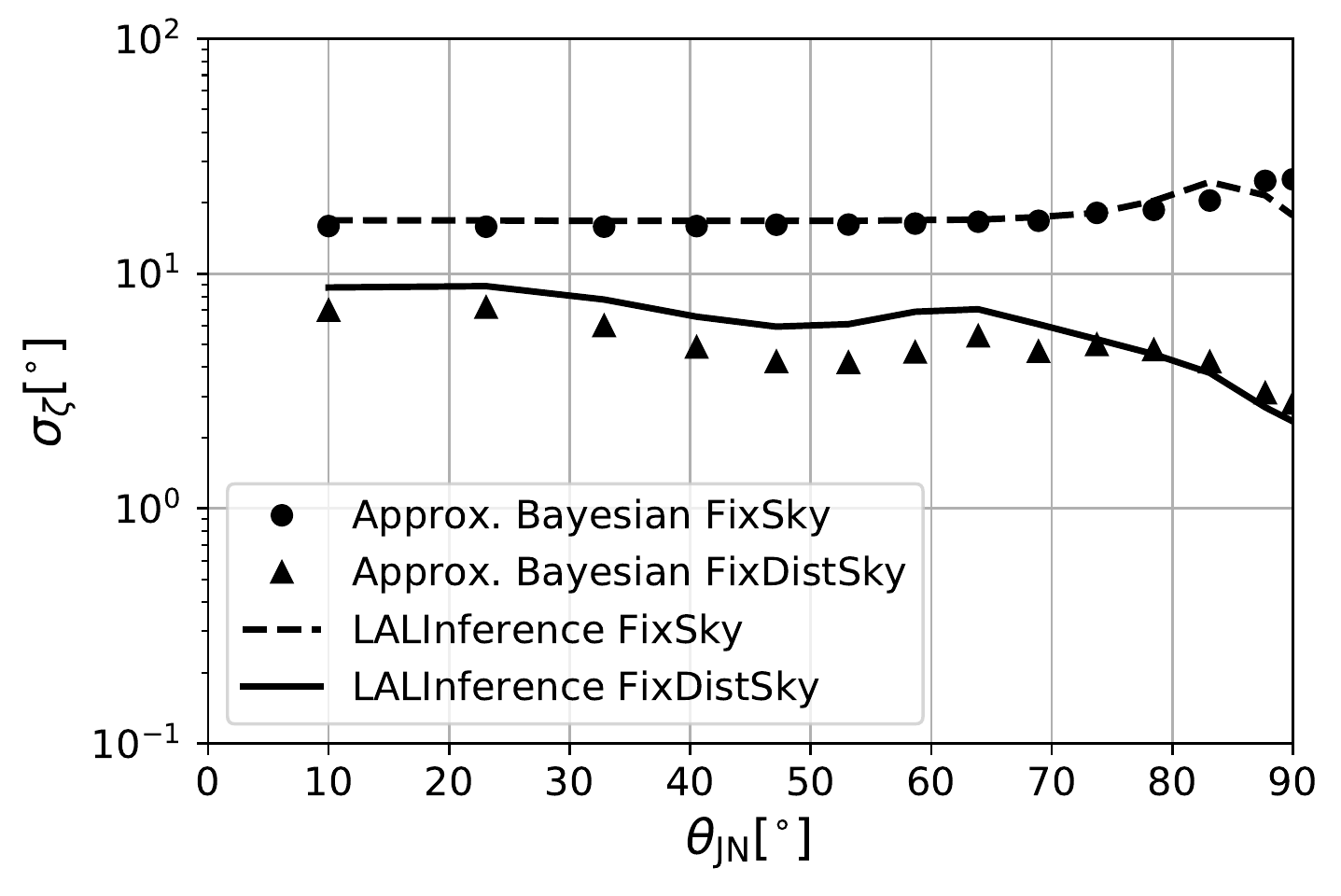}
\caption{\label{Fig.Comp}
Standard deviation for the viewing angle of source \weakevent against 
the true value of the inclination angle as measured with the full \linf code (lines) and with the approximate Bayesian estimator (symbols).
Triangles correspond to the case when both distance and sky position are known, while circles assume that only the sky position is known.}
\end{figure}

\subsection{Population analysis}

Having shown that the approximate Bayesian estimator gives results which are consistent with \linf, we proceed and use the former for large populations of BNSs.
We consider three observing scenarios~\citep{2018LRR....21....3A}: (i) A network with the two LIGOs and Virgo (HLV) 
at the expected LIGO-Virgo third observing run (O3) sensitivity~($\sim2019+$); (ii) HLV at the design sensitivity~($\sim2021+$).
(iii) To check how the results  evolve as the network of ground-based detectors grows, 
we also consider a five-detector network made of the two LIGOs, Virgo, LIGO India~\citep{ligoindia} and Kagra~\citep{2012CQGra..29l4007S,2013PhRvD..88d3007A} (HLVJI), all at their design sensitivity~($\sim2024+$).
For each scenario we simulate 1.4-1.4\msun BNSs with random orientation, distributed uniformly in comoving volume.
Throughout this work, a standard $\Lambda$-CDM Planck cosmology is assumed: $\Omega_{M_0}=0.3065,\Omega_{\Lambda_0}=0.6935,h_0=0.679$~\citep{2016A&A...594A..13P}.
A BNS is considered detected if the measured network signal-to-noise ratio~\footnote{Root-sum-square of 
individual detector's signal-to-noise ratio.}~\citep{2017arXiv170908079C} is greater than 12. 
Following the approach of Ref.~\cite{2015arXiv150900055C}, we add Gaussian noise to the measured SNR ratio, {relative arrival times} and relative phases.

We estimate the distance and inclination for 1000 detections, for each network.

\begin{itemize}

\item\textbf{Measurement of the viewing angle:} {We first discuss the measurability of the \va using auxiliary EM information (sky position and redshift).}

For the simulations in which we assume redshift information exists, we convert the redshift to \dl using the Planck cosmological parameters, and 
marginalize the posterior over the \dl with a Dirac $\delta$ centered at the true distance. 
We note that in practice this conversion might suffer from two sources of uncertainty: 
the redshift may not be a true measure of the luminosity distance due to the peculiar motion of the source, and the value of the Hubble constant is not precisely known.
{To account for the former, we introduce a $250$~km/s Gaussian uncertainty around the true value of the source redshift, which represents a typical 
uncertainty after the group velocity is corrected~\citep{2015MNRAS.450..317C,2018ApJ...859..101S}.} We also use a top-hat prior on the Hubble constant, from $65$ to $75$~km/s/Mpc to 
cover the range of currently estimated values~\citep{2016A&A...594A..13P,2016ApJ...826...56R}. 
Other cosmological parameters don't play a significant role for the redshift conversion, given that advanced detectors will only detect BNS up to redshift of $z<0.1$.

\begin{figure}
\centering
\includegraphics[width=1\columnwidth]{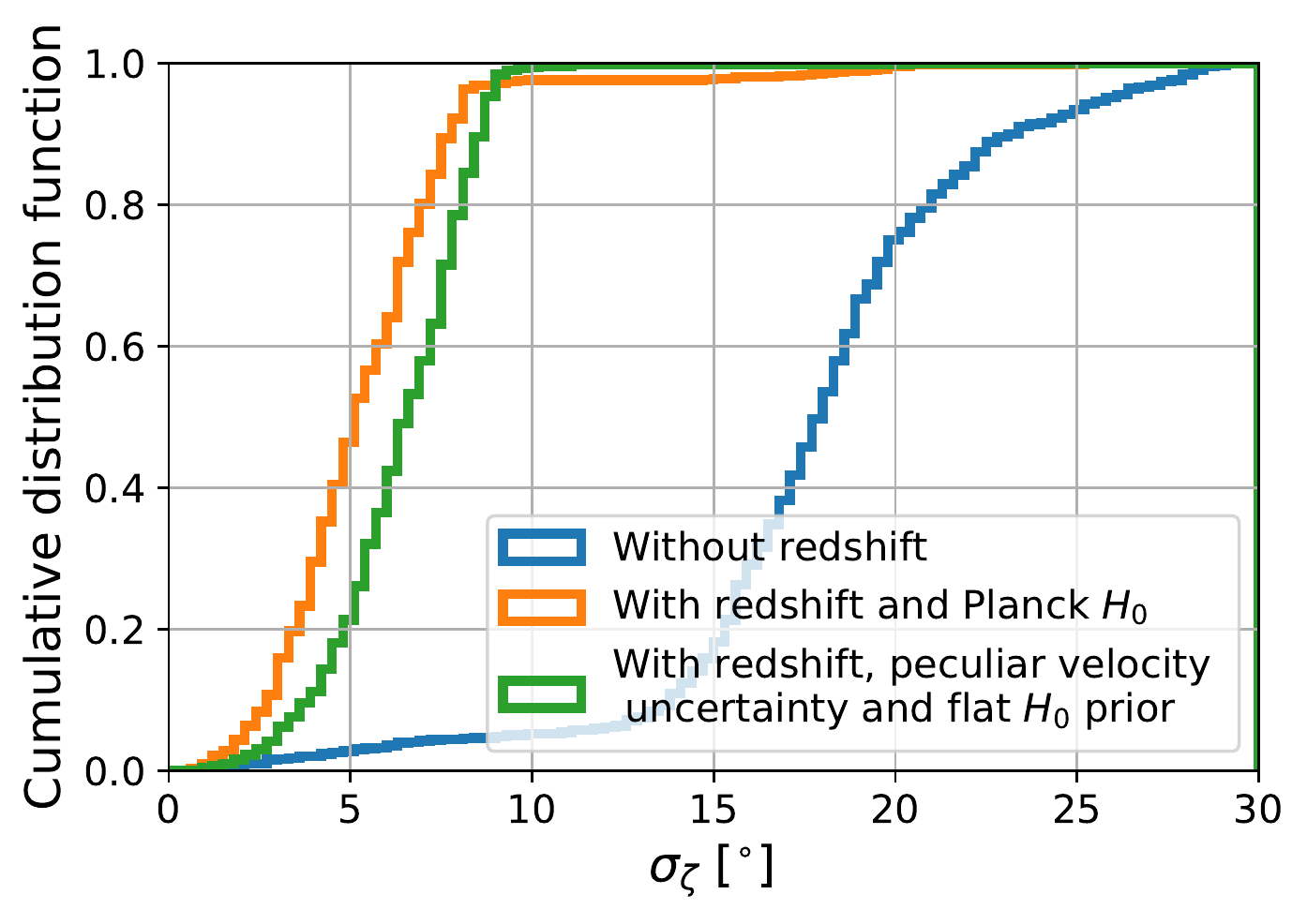}
\caption{\label{fig:cdfinc}
Cumulative distribution function of 1$\sigma$ \va  uncertainty for BNS detected by Advanced LIGO-Virgo at design sensitivity. 
{The sky positions for all events in the distributions are assumed to be precisely known.}}
\end{figure}
\bigskip
In Figure~\ref{fig:cdfinc} we show the cumulative distribution for the 1$\sigma$ \va uncertainty, 
for the HLV network at design sensitivity.
We see that if the sky positions and redshifts of the BNSs were precisely known, the \va for half of 
them would be constrained to $<6^{\circ}$. 
This uncertainty increases by $\sim 1^{\circ}$ if we include uncertainties in 
the peculiar velocity and in the cosmology. 

If instead only sky positions are identified, without any information about the \dl, only $<6\%$ of BNSs will have inclination 
uncertainty below $10^\circ$. 
These numbers are {comparable} for O3, and $\sim 10\%$ better when the detector network extends to HLVJI.

\bigskip
\item\textbf{Measurement of the luminosity distance:} Next, we wish to explore the situation where the EM sector provides a measurement or bound on the \va, and show how that can be used to measure more precisely the \dl of the binary using GW data. 
We consider various possibilities for the quality of the EM-based $\zeta$ measurement. The ideal scenario, in which the \va is \emph{precisely} measured; an uncertain measurement; and an upper bound.
In the first case, we use a Dirac $\delta$ centered at the true \va as prior for the inclination angle.
If an uncertain measurement is available we treat the $\zeta$ prior in the GW analysis as a normal distribution centered at the {true} value, 
and consider different widths of the distribution. 
{Last, we consider the case that EM sector only provide an upper bound on the \va $\zeta \leq \tjet$. 
This scenario can describe a situation in which a clear jet-break is observed in the GRB afterglow and the GRB jet 
is aligned with the binary axis, or any other measurement that might provide an upper limit on the \va.}

In Figure~\ref{fig:cdfdist} we show the cumulative distribution of the 1$\sigma$ fractional \dl uncertainty for simulated 
events detected by HLV at the design sensitivity. 
The blue curve shows the results for the worst case scenario, when no \va information is available. 
The green curves are obtained by assuming an uncertain Gaussian measurement from the EM sector, with standard deviation given in the legend.
Finally, the orange line is the optimal situation in which the \va is perfectly known. In this case, the bound on the \dl would be improved by a factor of $\sim 3$ compared to the worst case scenario.
\begin{figure}
\centering
\includegraphics[width=1\columnwidth]{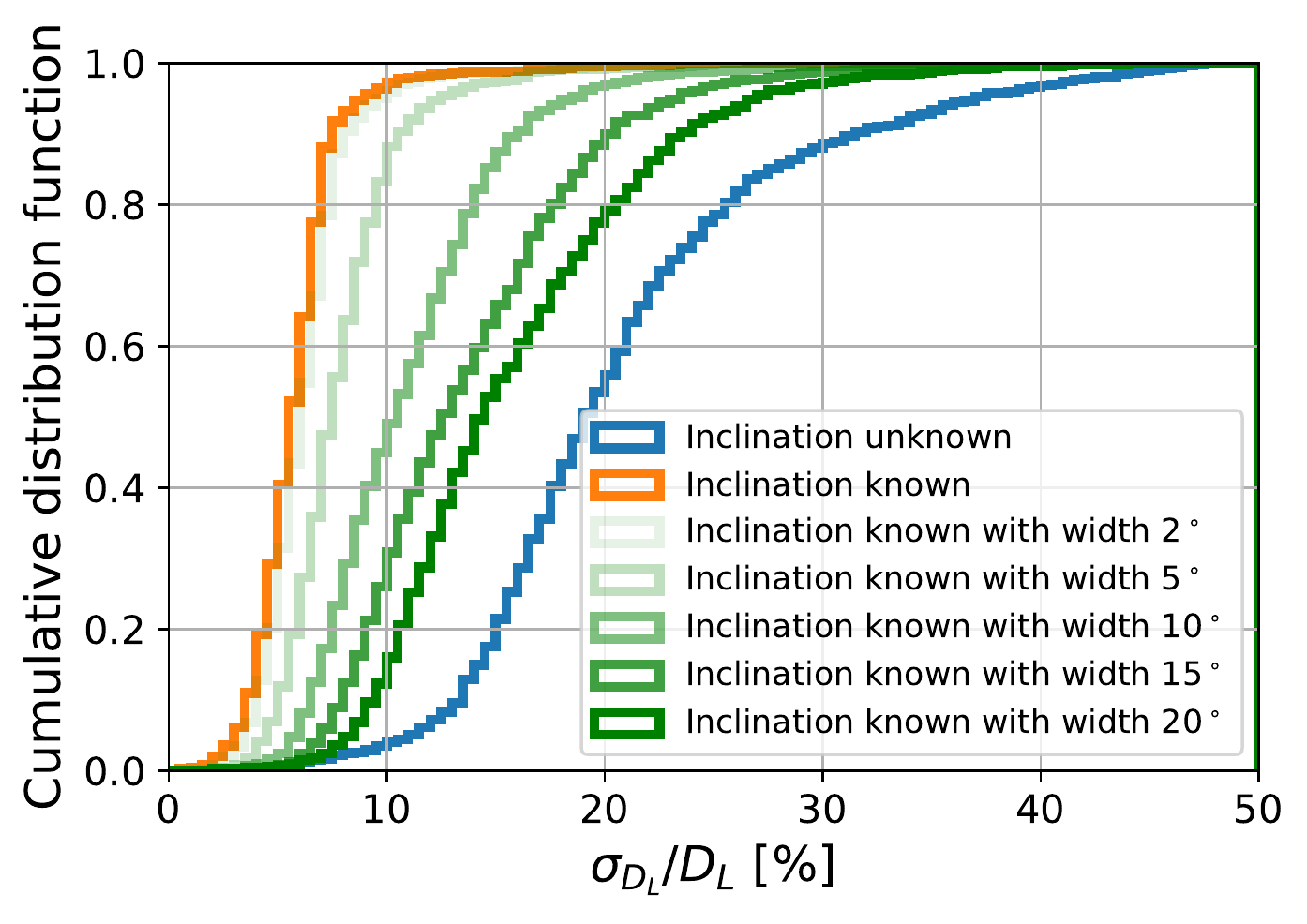}
\caption{\label{fig:cdfdist}
Cumulative distribution function of fractional \dl uncertainty, which is calculated as the 1$\sigma$ uncertainty divided by 
the true value, for BNS detected by Advanced LIGO-Virgo at design sensitivity.}
\end{figure}

{In Figure~\ref{fig:jet} we show results for the case when an upper bound \tjet is available for the \va.}
We consider various combinations of $(\tjn,\tjet)$, and for each pair we simulate 1000 detections.
We show median fractional \dl uncertainty for the 1000 BNS having the value of \tjn and \tjet given in the x axis and in the legend.
We find that, as long as $\tjet<30^{\circ}$, the fractional \dl uncertainty is $5\%$ to $7\%$ (depending on \tjn).

\begin{figure}
\centering
\includegraphics[width=1\columnwidth]{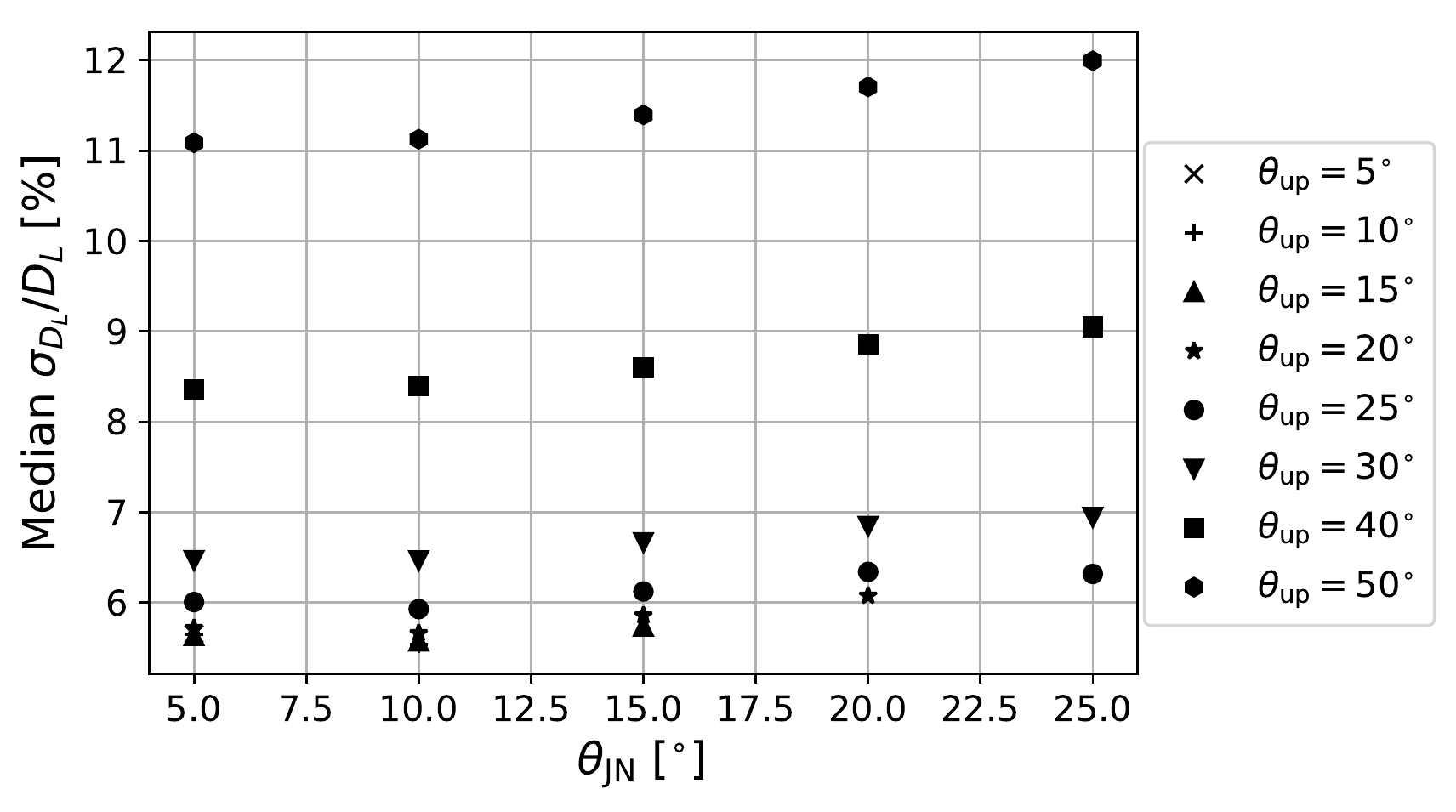}
\caption{\label{fig:jet}
{Median value of the fractional \dl uncertainty assuming the \va is bounded by the EM observations of BNSs detected by Advanced LIGO-Virgo at design sensitivity.
Each point represents the median uncertainty of 1000 BNSs with inclination angle $\tjn$ and EM upper-bound of \tjet.}
}
\end{figure}
\end{itemize}
\section{Discussion}

In this paper we have shown that $50\%$ of the BNSs detected by Advanced LIGO+Virgo for which sky position and redshift 
are known (from EM observations) will yield a 1$\sigma$ uncertainty on the \va of $7^{\circ}$ or less (see Figure~\ref{fig:cdfinc}). 
We emphasize that this result is \emph{independent} of any EM emission model. 
It therefore implies that the \va measurements obtained this way can be used to constrain the EM mechanism. 
The sky location and redshift of BNS can be solely determined  by the EM counterparts or, 
in some spectacular events, solely by the host~\citep{2016arXiv161201471C}--this is because in some well localized events the host groups can 
be uniquely identified without the help of EM counterparts. 

We have shown that BNSs without independent redshift information yield uninteresting inclination 
constraints unless the system is close to edge-on (see Figures~\ref{fig:incinc} and ~\ref{fig:cdfdist}). 
This is because the distance-inclination degeneracy dominates the uncertainty. Without an independent estimate of the redshift (and hence of the \dl), 
the degeneracy can only be broken when the binary is close to edge-on. 
Unfortunately, edge-on binaries are harder to detect and localize than face-on binaries; only $\sim 3\%$ of events will have \va$>80^\circ$. 
Even after advanced detectors start observing one BNS a week, only one or two events per year will be close to edge-on.
And yet, there are several reasons why binaries with large orbital inclinations should be sought out, beside yielding better 
inclination and \dl measurements (Figure~\ref{fig:incinc} and Ref.~\cite{2018arXiv180407337V}). 
For example, simulations suggest there may be interesting EM features along the equatorial plane~\cite{2015MNRAS.450.1777K}. 
EM follow-up observations will have to be properly planned in order to find the counterparts for these rare but valuable sources.  

{Adding two detectors to the network only improves the uncertainty of the \va by $\sim 10\%$. This is because the uncertainty on the \va is not only limited by how well one can measure the two polarizations, but also by the very correlation between \dl and \va,  which is significant for sources with small viewing angles, which are the majority.
Since HLV  can already measure two polarization, and more detectors won't help break the \dl-\va degeneracy, the uncertainty won't decrease much with larger networks. The only way to improve the measurement is to reduce the \dl-\va degeneracy, which can be done if the system shows amplitude modulation~\cite{2018arXiv180407337V,2004PhRvD..70d2001V}, or if the distance can be constrained by other means, e.g. an EM counterpart.}

The jet break in short GRB afterglows has long been used to study the jet opening angle~\citep{2015ApJ...815..102F}. 
The uncertainties in the inferred opening angle are usually a few degrees. 
It is interesting to verify how well GWs can constrain the BNS's viewing angle for those sources for which a GRB could be detected.
To answer this question we selected a subset of the BNSs in Figure~\ref{fig:cdfinc}, only keeping the sources with $\tjn<25^{\circ}$.
We find that half of this subsample of sources has 1$\sigma$ \va uncertainty of $8^{\circ}$ or less, if their sky locations and redshifts are constrained. 
That level of precision would be comparable with the jet opening angle uncertainty inferred from the afterglow jet break, allowing for a compelling comparison between the two (under the assumption that the total orbital angular momentum aligns with the jet).

We have also shown how one can expect a factor of 2 to 3 improvement in the fractional \dl uncertainty, if the binary inclination angle is independently measured. 
As the \dl is the main source of uncertainty when measuring the Hubble constant with GWs~\citep{1986Natur.323..310S}, 
a better \dl measurement translates into a better \hnot measurement.
Since the \hnot uncertainty scales as $1/\sqrt{N}$~\citep{2010ApJ...725..496N,2013arXiv1307.2638N,2018Natur.562..545C}, where $N$ is the number of detections, a factor of 3 improvement in distance uncertainty implies a factor of 9 fewer events are required to achieve any given \hnot precision. 
Instead of the 200 BNS detections required, as estimated in Ref.~\cite{2018Natur.562..545C}, only $\mathrm{O}(10)$ BNSs would be required to reach a {statistical} uncertainty of $1\%$ in \hnot, if the binary inclination angles were independently constrained. 
{{However, we stress} that the Hubble constant measurement is subject to systematic errors originating 
from both GW and EM measurements. 
From GW measurements, the dominating systematics is likely to be the instrumental calibration error in amplitude, 
which can potentially lead to a systematic bias in distance estimates~\citep{Vitale:2011wu}. This error is currently around a percent level, and is likely to improve~\citep{2016RScI...87k4503K}. }
{On the other hand,  the accuracy of EM measured binary \va relies on modeling of the EM emission. As more EM observations are made, EM modeling will improve, and those uncertainties will go down. But at least in the near future, the {systematic} uncertainty in GW-EM cosmological measurements will likely be dominated by the EM data.}

In Figure~\ref{fig:jet} we have shown that the median value of the fractional \dl uncertainty for BNSs can be as small as 5 to 7$\%$ 
if the {upper-bound \tjet on the \va from EM observation} is smaller than $30^{\degc}$.
This result is consistent with what was found by Ref.~\cite{2017PhRvL.119r1102F}. 
{We further find that as long as the upper-bound is smaller than $30^{\degc}$, 
the improvement in \dl measurement doesn't depend strongly on how tight the bound is. Conversely, 
the} \dl uncertainty is not significantly improved if {the upper-bound} is larger than 30$^{\circ}$. 
This is because, as shown in the Single-event analysis Section, the \va posteriors are equal to the \sd for true \va smaller than $\sim 70^\circ$.
Since the \sd peaks around 30$^{\circ}$ (black dashed line in Figure~\ref{fig:priorCompSourceA}), an EM-based upper limit only helps if 
it bounds the viewing angle to be smaller than $\sim 30^\circ$. 
{Fortunately, whenever an upper-bound on the \va is provided by a jet-break observation, the inferred jet opening angle of the short GRB almost never 
exceeds 30$^{\circ}$~\citep{1999A&A...344..573R,2005A&A...436..273A,2005NCimC..28..607R,2006ApJ...653..462G,2006ApJ...653..468B,2011ApJ...732L...6R,2015ApJ...815..102F,2016ApJ...827..102T,2018ApJ...857..128J}.} 
We thus expect that observations of {short GRB jet-breaks} will significantly improve the \dl measurement for BNSs. 

\acknowledgments

We acknowledge valuable discussions with K G Arun, Juan Calderon Bustillo, Maria Haney, Daniel Holz, Vivien Raymond, Om Sharan Salafia, B.S. Sathyaprakash, and John Veitch. 
H.-Y.C. was supported by the Black Hole Initiative at Harvard University, through a grant from the John Templeton Foundation.
S.V.~acknowledges support of the MIT physics department through the Solomon Buchsbaum Research Fund, the National Science Foundation, and the LIGO Laboratory. 
LIGO was constructed by the California Institute of Technology and Massachusetts Institute of Technology with funding from the National Science Foundation and operates under cooperative agreement PHY-0757058.
R.N. was supported in part by NSF grant AST1816420. The authors acknowledge the LIGO Data Grid clusters. We are grateful for computational resources provided by Cardiff University, and
funded by an STFC grant supporting UK Involvement in the Operation of Advanced LIGO.
We acknowledge useful comments from the anonymous referees.
LIGO Document Number P1800196.

\appendix

\section{Appendix}\label{sec:app}
\subsection{\linf inclination angle posteriors}

In Figures.~\ref{fig:priorCompSourceA}, \ref{fig:priorCompSourceC}, \ref{fig:priorCompSourceD}, \ref{fig:priorCompSourceB} we show the posteriors on the viewing angle obtained with \linf when no EM information is provided. 
Thin lines (all overlapping) correspond to small and moderate inclination angles. We use thicker lines for sources with$\tjn=74^\degc, 78^\circ, 83^\circ, 88^\circ, 90^\circ$. 
For these sources the viewing angle posterior is usually significantly different from the prior. 
We stress that source \loudevent is quite loud, with a network SNR of 35, similar to GW170817~\cite{2017PhRvL.119p1101A}.
For the weakest event, source \weakevent, the SNR in so low that the two polarizations cannot be disentangled even partially, 
and the posteriors are exactly equal to the \sd, with an uncertainty of $19.5^\degc$ for small and moderate inclinations 
(the standard deviation on the viewing angle of the \sd, inferred from the analytic expression in Ref.~\cite{2011CQGra..28l5023S}, is $19.4^\degc$).
\begin{figure}
\includegraphics[width=1\columnwidth]{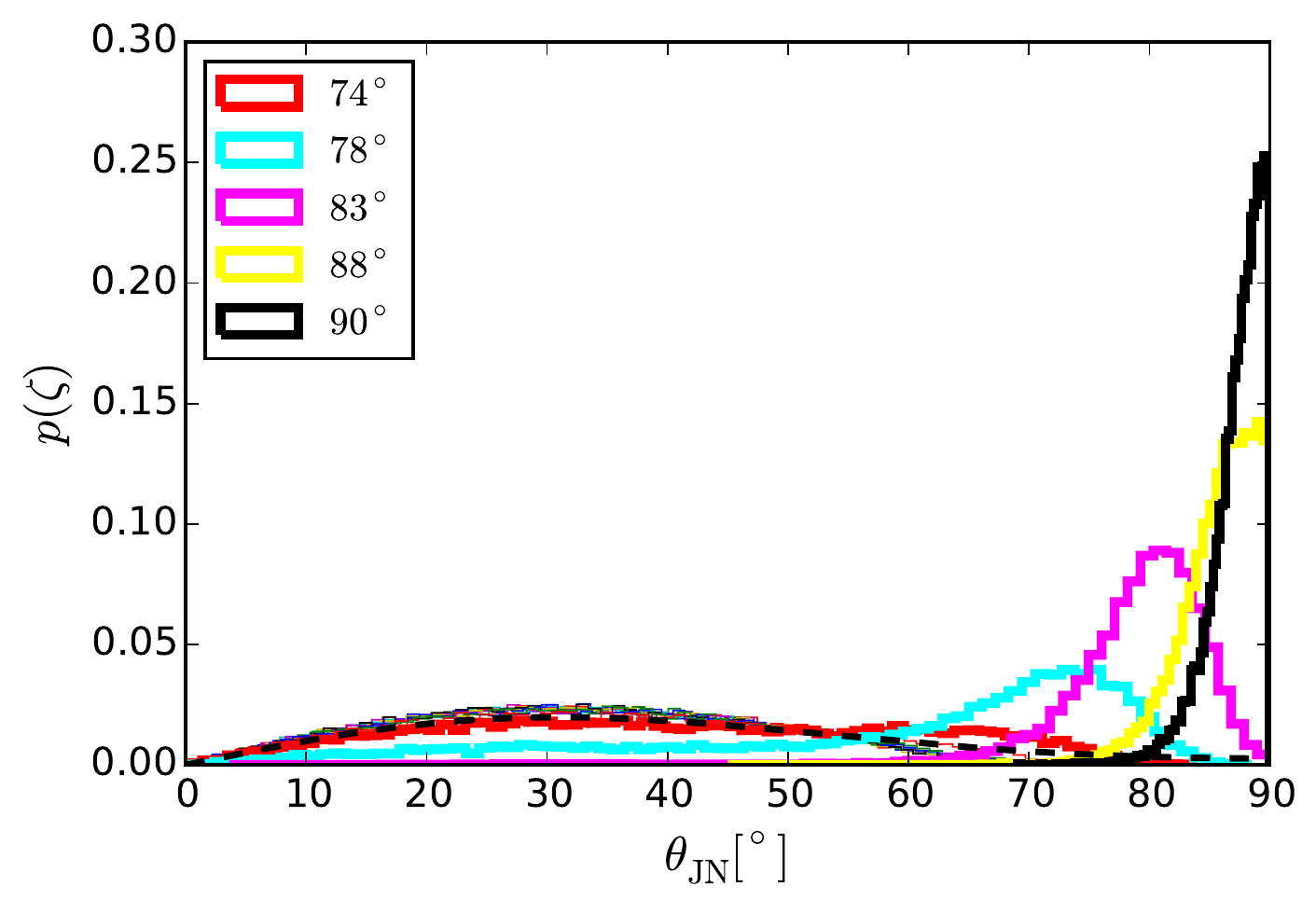}
\caption{Same as Figure~\ref{fig:priorCompSourceA}, but for source \loudevent.}\label{fig:priorCompSourceC}
\end{figure}

\begin{figure}
\includegraphics[width=1\columnwidth]{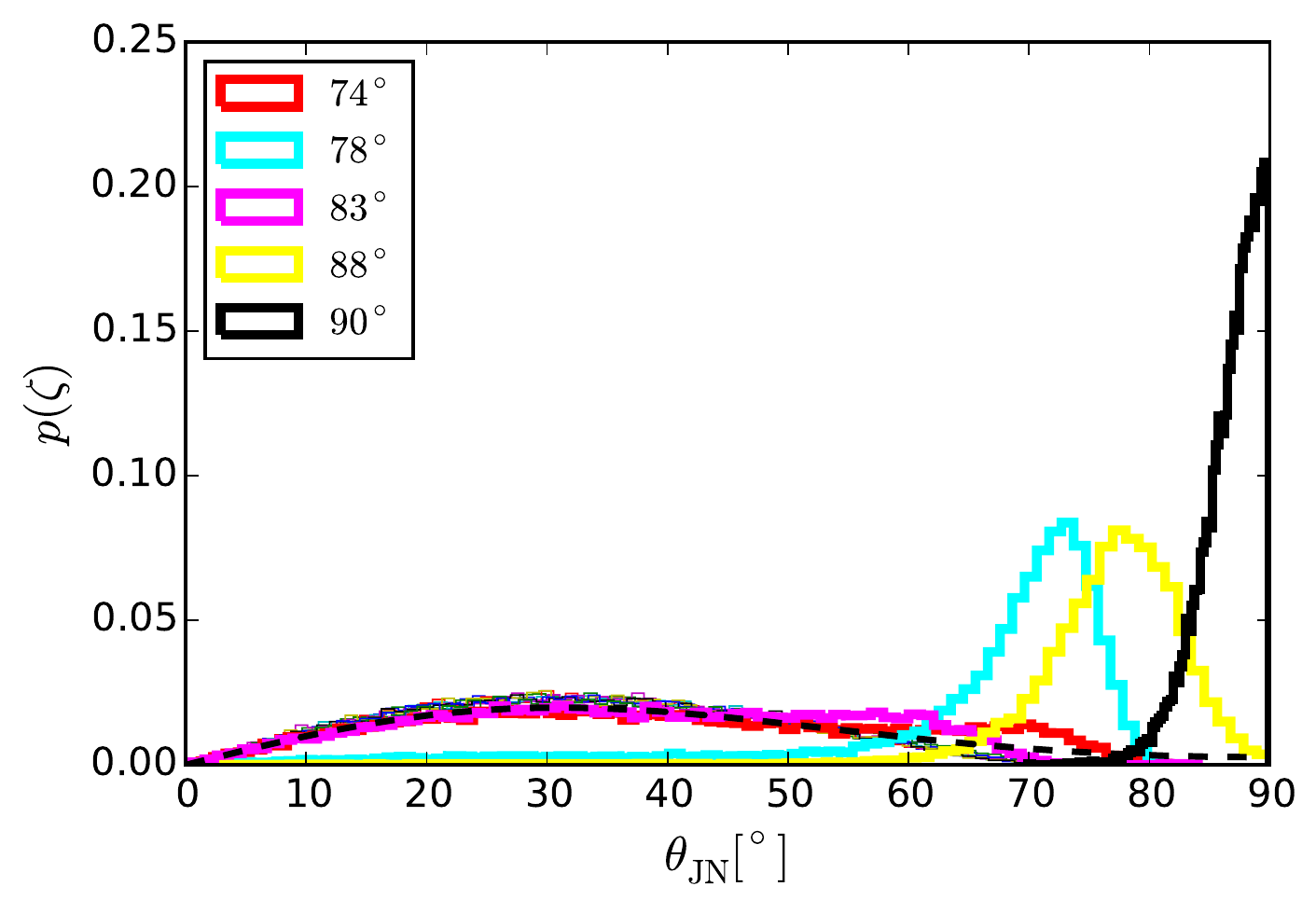}
\caption{Same as Figure~\ref{fig:priorCompSourceA}, but for source \polarevent.}\label{fig:priorCompSourceD}
\end{figure}

\begin{figure}
\includegraphics[width=1\columnwidth]{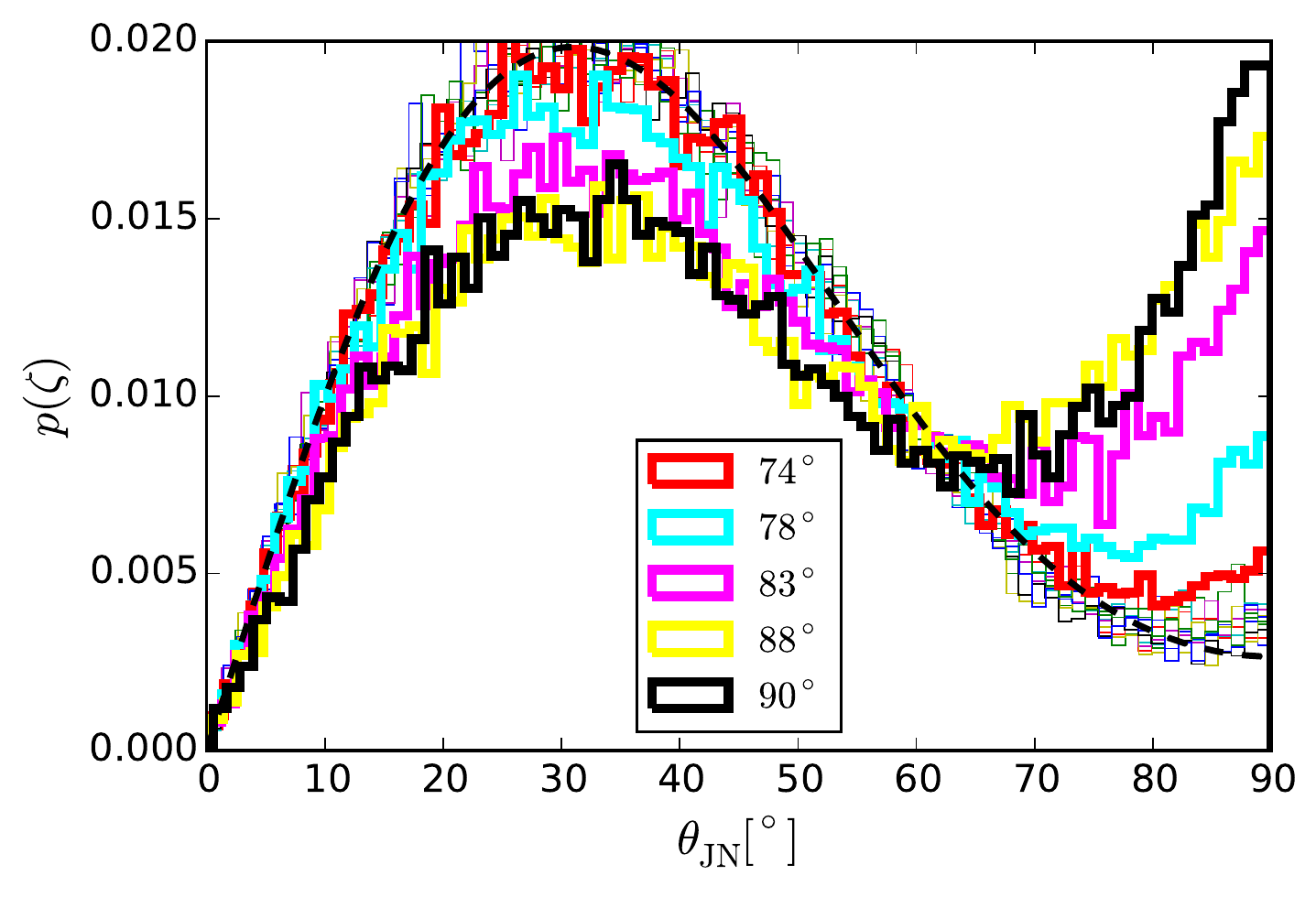}
\caption{Same as Figure~\ref{fig:priorCompSourceA}, but for source \weakevent.}\label{fig:priorCompSourceB}
\end{figure}

\subsection{Approximate Bayesian estimator}
We use each individual detector's signal-to-noise ratio, $\vec{\rho}=(\rho_a,\rho_b,\rho_c...)$ [a, b, c etc. denote 
different detectors], {the relative arrival time difference between detector pairs, 
$\vec{\Delta t}=(\Delta t_{ab},\Delta t_{ac},...)$,}~\footnote{{$\Delta t_{bc}$ can be obtained as the difference of $\Delta t_{ab}$ and $\Delta t_{ac}$}} and the relative phase 
difference between detector pairs, $\vec{\Delta \eta}=(\Delta \eta_{ab},\Delta \eta_{ac},...)$, to 
reconstruct the distance $D$ and inclination $\tjn$ posterior:
\begin{align}\label{eq:posterior}
\nonumber &f(D,\tjn|\vec{\rho},\vec{\Delta t},\vec{\Delta \eta}) \propto f(D,\tjn)f(\vec{\rho},\vec{\Delta t},\vec{\Delta \eta}|D,\tjn) \\
\nonumber &= f(D,\tjn) \frac{1}{f(D,\tjn)} \displaystyle\int f(\vec{\rho},\vec{\Delta t},\vec{\Delta \eta}|D,\tjn,\psi) f(D,\tjn,\psi) d\psi \\
&= \displaystyle\int f(\vec{\rho},\vec{\Delta t},\vec{\Delta \eta}|D,\tjn,\psi) f(D,\tjn,\psi) d\psi.
\end{align} 
The first line of Equation~\ref{eq:posterior} follows the Bayes theorem. 
$f(D,\tjn)$ is the prior on the distance and the binary inclination angle, and $f(\vec{\rho},\vec{\Delta t},\vec{\Delta \eta}|D,\tjn)$
is the likelihood. Since the signal-to-noise ratio and the phase difference also depend on the binary orientation angle $\psi$ 
(we initially don't write the binary orientation out explicitly because we are not interested in this quantity), we write 
the likelihood as a marginalization over the orientation angle in the second line of Equation~\ref{eq:posterior}. 
We then need a different prior $f(D,\tjn,\psi)$. This prior can be written as 
\begin{equation}\label{eq:prior}
f(D,\tjn,\psi)=D^2\, \rm{sin}\,\tjn\, H[D_h(\tjn,\psi)- D]
\end{equation}
where $H[D_h(\tjn,\psi)- D]$ is a Heaviside function that cuts off at the maximum distance $D_h(\tjn,\psi)$ a binary 
with inclination and orientation $(\tjn,\psi)$ can be detected. The new likelihood $f(\vec{\rho},\vec{\Delta t},\vec{\Delta \eta}|D,\tjn,\psi)$ is calculated 
as $f(\vec{\rho},\vec{\Delta t},\vec{\Delta \eta}|D,\tjn,\psi)\sim {\rm exp}(-\chi^2_{\rho}/2)\,{\rm exp}(-\chi^2_{\zeta}/2)$, 
where
\begin{equation}\label{eq:chisnr}
\chi^2_{\rho}=\frac{\displaystyle\sum_i(\rho_{{\rm measured},i}-\rho_i (D,\tjn,\psi))^2}{\sigma_{\rho}^2}.
\end{equation}

where the index $i$ goes through all detectors. 

$\sigma_{\rho}^2$ and $\chi^2_{\zeta}$ are taken from Equations 10 to 14 of Ref.~\cite{2015arXiv150900055C}. 
We grid the $(D,\tjn,\psi)$ parameter space and evaluate $\chi^2_{\rho}$, $\chi^2_{\zeta}$ and $f(D,\tjn,\psi)$ in each grid cell. These terms are thus numerically integrated over $\psi$ to yield the joint distance-inclination posterior, Equation~\ref{eq:posterior}.

\bibliography{ref}

\end{document}